\documentclass[prd,showpacs,12pt]{revtex4}
\usepackage{bm}
\usepackage{graphics}
\usepackage{rotating}
\usepackage{epsfig}

\begin{document} 
\begin{flushright}{\rm\normalsize HU-EP-02/19}\end{flushright}
\title{Charm mass corrections to the bottomonium mass spectrum}
\author{D. Ebert}
\affiliation{Institut f\"ur Physik, Humboldt--Universit\"at zu Berlin,
Invalidenstr.110, D-10115 Berlin, Germany}
\author{R. N. Faustov}
\author{V. O. Galkin}
\affiliation{Institut f\"ur Physik, Humboldt--Universit\"at zu Berlin,
Invalidenstr.110, D-10115 Berlin, Germany}
\affiliation{Russian Academy of Sciences, Scientific Council for
Cybernetics, Vavilov Street 40, Moscow 117333, Russia}

\begin{abstract}
The one-loop corrections to the bottomonium mass spectrum due to the
finite charm mass are evaluated in the framework of the relativistic
quark model. The obtained corrections are compared with the results of
perturbative QCD.  
\end{abstract}
\maketitle

Recently effects of the finite mass of the charm quark in
determining the bottomonium mass spectrum and the $b$ quark mass were
considered within perturbative QCD \cite{hm,h,m,bsv}. Here we study
these effects in the framework of the relativistic quark model taking
into account the confining quark interaction. Since the $b\bar b$
bound system is rather far from being a Coulombic one (even in the
ground state) we expect the size of these effects to differ noticeably
from that predicted within perturbative QCD, where they were found to
be substantial \cite{bsv}. This report may be
considered as completing our previous paper \cite{efg}, where the mass
spectra of heavy quarkonia were calculated with the account of the
one-loop radiative corrections. Our model was also successfully
applied for describing the heavy-light meson mass spectra  and
radiative and  weak decays of heavy mesons \cite{gf,egf}.

In papers  \cite{egf,efg,gf} a meson is described by the wave
function of the bound quark-antiquark state, which satisfies the
quasipotential equation of the Schr\"odinger type in
the center-of-mass frame:
\begin{equation}
\label{quas}
{\left(\frac{b^2(M)}{2\mu_{R}}-\frac{{\bf
p}^2}{2\mu_{R}}\right)\Psi_{M}({\bf p})} =\int\frac{d^3 q}{(2\pi)^3}
 V({\bf p,q};M)\Psi_{M}({\bf q}),
\end{equation}
where the relativistic reduced mass
is
\begin{equation}\label{mur}
\mu_{R}=\frac{M^4-(m^2_1-m^2_2)^2}{4M^3},
\end{equation}
and 
$b^2(M)$  denotes
the on-mass-shell relative momentum squared
\begin{equation}
\label{bm}
{b^2(M) }
=\frac{[M^2-(m_1+m_2)^2][M^2-(m_1-m_2)^2]}{4M^2}.
\end{equation}
Here $m_{1,2}$ and $M$ are quark masses
and a meson mass, respectively. For the $b\bar b$ bound system
(bottomonium) $m_1=m_2=m_b$ and Eqs. (\ref{mur}), (\ref{bm}) take
the form
\begin{equation}
  \label{eq:bb}
  \mu_R=\frac{M}{4}, \qquad b^2(M)=\frac{M^2}{4}-m_b^2.
\end{equation}

The kernel
$V({\bf p,q};M)$ in Eq.~(\ref{quas}) is the quasipotential operator of
the quark-antiquark interaction. It is constructed with the help of the
off-mass-shell scattering amplitude, projected onto the positive
energy states. An important role in this construction is played
by the Lorentz-structure of the confining quark-antiquark interaction
in the meson.  In
constructing the quasipotential of the quark-antiquark interaction
we have assumed that the effective
interaction is the sum of the usual one-gluon exchange term and the mixture
of vector and scalar linear confining potentials.
The quasipotential is then defined by
\cite{efg}
\begin{equation}
\label{qpot}
V({\bf p,q};M)=\bar{u}_1(p)\bar{u}_2(-p){\mathcal V}({\bf p}, {\bf
q};M)u_1(q)u_2(-q),
\end{equation}
with
\begin{equation}\label{vcal}
{\mathcal V}({\bf p},{\bf q};M)=\frac{4}{3}\alpha_sD_{ \mu\nu}({\bf
k})\gamma_1^{\mu}\gamma_2^{\nu}
+V^V_{\rm conf}({\bf k})\Gamma_1^{\mu}
\Gamma_{2;\mu}+V^S_{\rm conf}({\bf k}),
\end{equation}
where $\alpha_s$ is the QCD coupling constant, $D_{\mu\nu}$ is the
gluon propagator in the Coulomb gauge
and ${\bf k=p-q}$; $\gamma_{\mu}$ and $u(p)$ are
the Dirac matrices and spinors.
The effective long-range vector vertex is
given by
\begin{equation}
\Gamma_{\mu}({\bf k})=\gamma_{\mu}+
\frac{i\kappa}{2m_b}\sigma_{\mu\nu}k^{\nu}, \qquad k^\nu=(0,{\bf k}), 
\end{equation}
where $\kappa$ is the Pauli interaction constant characterizing the
nonperturbative anomalous chromomagnetic moment of quarks. Vector and
scalar confining potentials in the nonrelativistic limit reduce to
\begin{equation}\label{vconf}
V^V_{\rm conf}(r)=(1-\varepsilon)Ar+B,\qquad
V^S_{\rm conf}(r) =\varepsilon\, Ar,
\end{equation}
reproducing
\begin{equation}
V_{\rm conf}(r)=V^S_{\rm conf}(r)+
V^V_{\rm conf}(r)=Ar+B,
\end{equation}
where $\varepsilon$ is the mixing coefficient.

The quasipotential for the heavy quarkonia, including retardation,
one-loop radiative corrections and
expanded in $p^2/m^2$, can be found in Ref.~\cite{efg}.
All the parameters of
our model, such as quark masses, parameters of the linear confining potential,
mixing coefficient $\varepsilon$ and anomalous
chromomagnetic quark moment $\kappa$, were fixed from the analysis of
heavy quarkonium spectra \cite{efg} and radiative decays \cite{gf}.
The quark masses
$m_b=4.88$ GeV, $m_c=1.55$ GeV, $m_s=0.50$ GeV, $m_{u,d}=0.33$ GeV and
the slope of the linear potential $A=0.18$ GeV$^2$ in our model have
the fixed values which agree with the usually accepted. The constant
term $B=-0.16$ GeV may be adjusted producing the overall level shift.
In Ref.~\cite{fg} we have considered the expansion of  the matrix
elements of weak heavy quark currents between pseudoscalar and vector
meson ground states up to the second order in inverse powers of
the heavy quark
masses. It has been found that the general structure of the leading,
first,
and second order $1/m_Q$ corrections in our relativistic model is in accord
with the predictions of heavy quark effective theory. The heavy quark
symmetry and QCD impose rigid 
constraints on the parameters of the long-range potential in our model.
The analysis of the first order corrections  fixes the value
of the Pauli interaction constant $\kappa=-1$ \cite{fg}. The same
value of $\kappa$  was found previously
from  the fine splitting of heavy quarkonia ${}^3P_J$- states \cite{efg}.
The value of the parameter characterizing the  mixing of
vector and scalar confining potentials, $\varepsilon=-1$,
was found from the analysis of the $1/m_Q^2$ corrections \cite{fg}
and from considering radiative decays of heavy quarkonia \cite{gf}.

The perturbative heavy quark potential to two loops including the
effects of massive loop quarks can be found in Ref.~\cite{m}. In our
model we take into account only one-loop contributions since the
uncertainty brought by the confining interaction is larger then the
two-loop contributions. The one-loop correction with the finite $c$
quark mass to the Coulomb potential can be included in the form
\cite{m}:
\begin{eqnarray}
  \label{eq:cp}
  V(r,m_c)&=&-C_F\frac{\alpha_V(r,m_c)}r,\cr\cr
\alpha_V(r,m)&=&\alpha_s(\mu)\left[1+v_1(r,m,\mu)\frac{\alpha_s(\mu)}{\pi}
\right], \cr\cr
v_1(r,m,\mu)&=&\frac{C_A}4\left\{\frac{31}9+\frac{22}3\left[\ln(\mu r)+
    \gamma_E\right] \right\}-\frac59 T_F+\frac{T_F}3\left[\ln\frac{a_0
    m^2}{\mu^2}+2E_1(\sqrt{a_0}mr)\right],
\end{eqnarray}
where
\[
E_1(x)=\int_x^\infty e^{-t}\ \frac{dt}t=-\gamma_E-\ln x-
\sum_{n=1}^\infty \frac{(-x)^n}{n\cdot n!},
\] 
the Euler constant $\gamma_E\cong0.5772$ and $C_A=3$, $C_F=4/3$,
$T_F=1/2$ in QCD. A simple representation for the vacuum polarization
operator with $a_0=5.2$ was used \cite{m}. Subtracting from
$v_1(r,m,\mu)$ its value at $m=0$, namely
\[
v_1(r,0,\mu)=\frac{31}{36}C_A-\frac59 T_F+\left(\frac{11}6 C_A
  -\frac23 T_F\right)\left[\ln(\mu r)+\gamma_E\right],
\]
we obtain
\begin{equation}
  \label{eq:dv}
  \Delta v_1(r,m,\mu)=v_1(r,m,\mu)-v_1(r,0,\mu)=\frac23 T_F\left[
  \ln(\sqrt{a_0}mr) +\gamma_E+E_1(\sqrt{a_0}mr)\right].
\end{equation}
Thus the one-loop correction to the static $Q\bar Q$ potential in QCD
due to the finite $c$ quark mass reads as:
\begin{equation}
  \label{eq:deltav}
  \Delta V(r,m_c)=-\frac43\frac{\alpha_s^2(\mu)}{\pi r}\Delta
  v_1(r,m_c,\mu)= -\frac{\alpha_s^2(\mu)}{\pi
  r}\frac49\left[\ln(\sqrt{a_0}m_cr) +\gamma_E+E_1(\sqrt{a_0}m_cr)\right].
\end{equation}
The charm mass correction to the bottomonium mass spectrum is then given by
\begin{equation}
  \label{eq:dm}
\Delta M=\left<\Delta V_{m_c}\right>, \qquad \alpha_s(m_b)=0.22, \qquad
m_c=1.55\ {\rm GeV}.
\end{equation}
For averaging we use both the wave functions obtained in calculating
the heavy quarkonium mass spectra \cite{efg} and the Coulomb wave
functions. The Coulomb averaging is carried out also for
$\alpha_s=0.3$ sometimes used in the perturbative QCD description of
bottomonium. The numerical values of this spin-independent correction
for different $b\bar b$ states are 
presented in the Table~\ref{tab:1}. 
\begin{table}[htbp]
  \centering
 \caption{Charm mass corrections to the masses of  bottomonium
   $\{n_r+1\}L(n=n_r+L+1)$  states (in MeV).}  
  \label{tab:1}
\begin{ruledtabular}
  \begin{tabular}{ccccccc}
State& $1S(1)$ & $1P(2)$& $2S(2)$ & $1D(3)$ &
$2P(3)$ & $3S(3)$\\ 
\hline
$\left<\Delta V_{m_c}\right>$ & $-12$&$-9.3$ &$-8.7$&
$-7.6$&$-7.5$&$-7.2$\\
$\left<\Delta V_{m_c}\right>_{\rm Coul}^{\alpha_s=0.22}$&$-9.5$ &$-4.2$
&$-3.8$ &$-2.3$ &$-2.2$ &$-2.1$ \\
$\left<\Delta V_{m_c}\right>_{\rm Coul}^{\alpha_s=0.3}$&$-20.7$ &$-9.7$
&$-8.8$ &$-5.5$ &$-5.2$ &$-4.9$ \\
\hline
$\left<\Delta V_{m_c}\right>_{\rm Coul}$ \cite{bsv} &$-14.3$ &$-22.1$
&$-21.9$ &  & $-49$ & $-40.5$\\
$\alpha_s(\mu)$ & 0.277 & 0.437 & 0.452 & & 0.733 & 0.698 
\end{tabular}  
\end{ruledtabular}
 \end{table}

The Table~\ref{tab:1} shows that for a fixed value of $\alpha_s$ the
averagings with and without confining potential substantially differ
especially for the excited states. For growing $n=n_r+L+1$ the values
of $\left<\Delta V_{m_c}\right>$ slowly decrease in our model whereas
for the Coulomb potential they fall rapidly. The growth of $(\delta E_{b\bar
  b})_{m_c}^{(1)}\equiv \left<\Delta V_{m_c}\right>_{\rm Coul}$ in
Ref.~\cite{bsv} (Table~I) is evoked by fast increasing values of
$\alpha_s(\mu)$. For 
these values of  $\alpha_s(\mu)$ we also reproduce all the corrections
$(\delta E_{b\bar b})_{m_c}^{(1)}$ obtained in \cite{bsv}. The values
of the scale $\mu$ were fixed in Ref.~\cite{bsv} from the stability
conditions containing the corrections $\delta m_b$ to the $b$ quark
mass which are absent in our model.  

We estimate the uncertainty in calculating the spin-averaged mass
spectrum of the bottomonium to be few MeV. Some of its sources are
the uncertainties in the quark masses and confining potential  as well
as higher order relativistic and radiative
corrections. The present knowledge of the spin-independent part of the heavy
quark potential  is rather uncertain since it strongly depends on the
Lorentz structure of the confining potential. The calculated
corrections partly may be absorbed in the definition of
the constant term in the static potential (\ref{vconf}). Nevertheless
they may become essential in future 
when the long-range $Q\bar Q$ interaction will be known better. Then
the heavy quark mass should be treated more self-consistently not as
a fixed phenomenological parameter but as a scale-dependent one
normalized from the heavy quarkonium mass spectrum.

The authors express their gratitude to  N. Brambilla, 
M. M\"uller-Preussker and V. Savrin  
for support and discussions. Two of us (R.N.F and V.O.G.)
were supported in part by the {\it Deutsche
Forschungsgemeinschaft} under contract Eb 139/2-1, {\it
 Russian Foundation for Fundamental Research} under Grant No.\
00-02-17768 and {\it Russian Ministry of Education} under Grant
No. E00-3.3-45.

\end{document}